\documentclass[%
 reprint,
 amsmath,amssymb,
prl,
floatfix,
]{revtex4-1}

\usepackage{graphicx}
\usepackage{dcolumn}
\usepackage{bm}
\usepackage{epstopdf}

\begin{document}

\title{Saddles, Twists, and Curls: Shape Transitions in Freestanding Nanoribbons}

\author{Hailong Wang}
\author{Moneesh Upmanyu}
\email{mupmanyu@neu.edu}
\affiliation{Group for Simulation and Theory of Atomic-Scale Material Phenomena (stAMP), Department of Mechanical and Industry Engineering, Northeastern University, Boston, Massachusetts 02115, USA.}

\begin{abstract}
Efforts to modulate the electronic properties of atomically thin crystalline nanoribbons requires precise control over their morphology. Here, we perform atomistic simulations on freestanding graphene nanoribbons (GNRs) to first identify the minimal shapes, and then employ a core-edge framework based on classical plate theory to quantify the width dependence in more general systems. The elastic edge-edge interactions force ultra-narrow ribbons to be flat, which then bifurcate to twisted and bent shapes at critical widths that vary inversely with edge stress. Compressive edge stresses results in twisted and saddle shapes that are energetically indistinguishable in the vicinity of the bifurcation. 
Increasing widths favor the saddle shapes with (longitudinal) ribbon curvatures that vary non-linearly with width and edge stress. Positive edge stresses result in a flat-to-curled transition with similar scalings. At large widths with negligible edge-edge interactions, rippling instabilities set in, i.e. edge ripples and midline dimples for compressive and tensile edge stresses. Our results highlight the utility of the core-edge framework in developing a unified understanding of the interplay between geometry and mechanics that sets the morphology of crystalline nanoribbons.



\end{abstract}
\maketitle
The electronic properties of crystalline nanoribbons are set by their atomic-scale structure, both within the ribbon core and at the edges. The properties are naturally modulated by the ribbon morphology and past computational studies on graphene nanoribbons (GNRs) have identified several shapes for varying widths $w$ and edge stresses $\tau_e$. For compressive edge stresses, semi-infinite sheets exhibit classical edge ripples~\cite{gph:ShenoyZhang:2008, gnr:LuHuang:2010} that spontaneously twist below a critical width~\cite{gnr:BetsYakobson:2009}. At even smaller widths ($w<1.5$\,nm), quantum computations show that the ribbon becomes flat~\cite{gnr:KitKoskinen:2010}. Little is known about the interplay between geometry and mechanics that drives these shape transitions. Intuitively, the accommodation of the relative stretch at the edge (or compression for $\tau_e>0$) sets the shape stability. More formally, the edge stresses induce a non-Euclidean metric as they relax, $ds^2=[g(y)\,dx]^2 + dy^2$, 
where $s$ is a surface arc length, $x$ and $y$ are coordinates along and normal to the ribbon midline and $g(y)$ is the reference (target) metric that captures the interplay; $g=1$ away from the edge and changes as we approach the edge in accordance with the sign of $\tau_e$. The target morphology represents an isometric embedding, i.e. no stretching~\cite{memb:SharonSwinney:2002, *memb:AudolyBoudaoud:2003}. The standard recipe is to seek a shape that minimizes the total energy and the metric incompatibility using plate theory which, for linear elastic thin sheets (thickness $h\ll l$, where $l$ is the extent of the sheet) and for small strains, reduces to the classical F\"oppl-van K\'arm\'an (F-vK) equations~\cite{book:Mansfield:1989}. 

The atomically thin systems represent the extreme limit $h\rightarrow0$ and one would naively expect the minimal energy surface to be almost inextensible. 
That is not strictly true for the nanoribbons considered here as they are inherently confined due to the edge-edge interactions. We therefore rely on global minimum energy shapes observed in high fidelity atomic-scale simulations to identify the competing morphologies. The shapes serve as inputs to a composite (core-edge) framework, based on the full F-vK equations, that is employed to explore the morphological space as a function of their geometry ($w$) and material parameters (bending and stretching stiffness $D$ and $S$, edge stress and stiffness, $S_e$).

\begin{figure}[tb]
\includegraphics[width=0.8 \columnwidth]{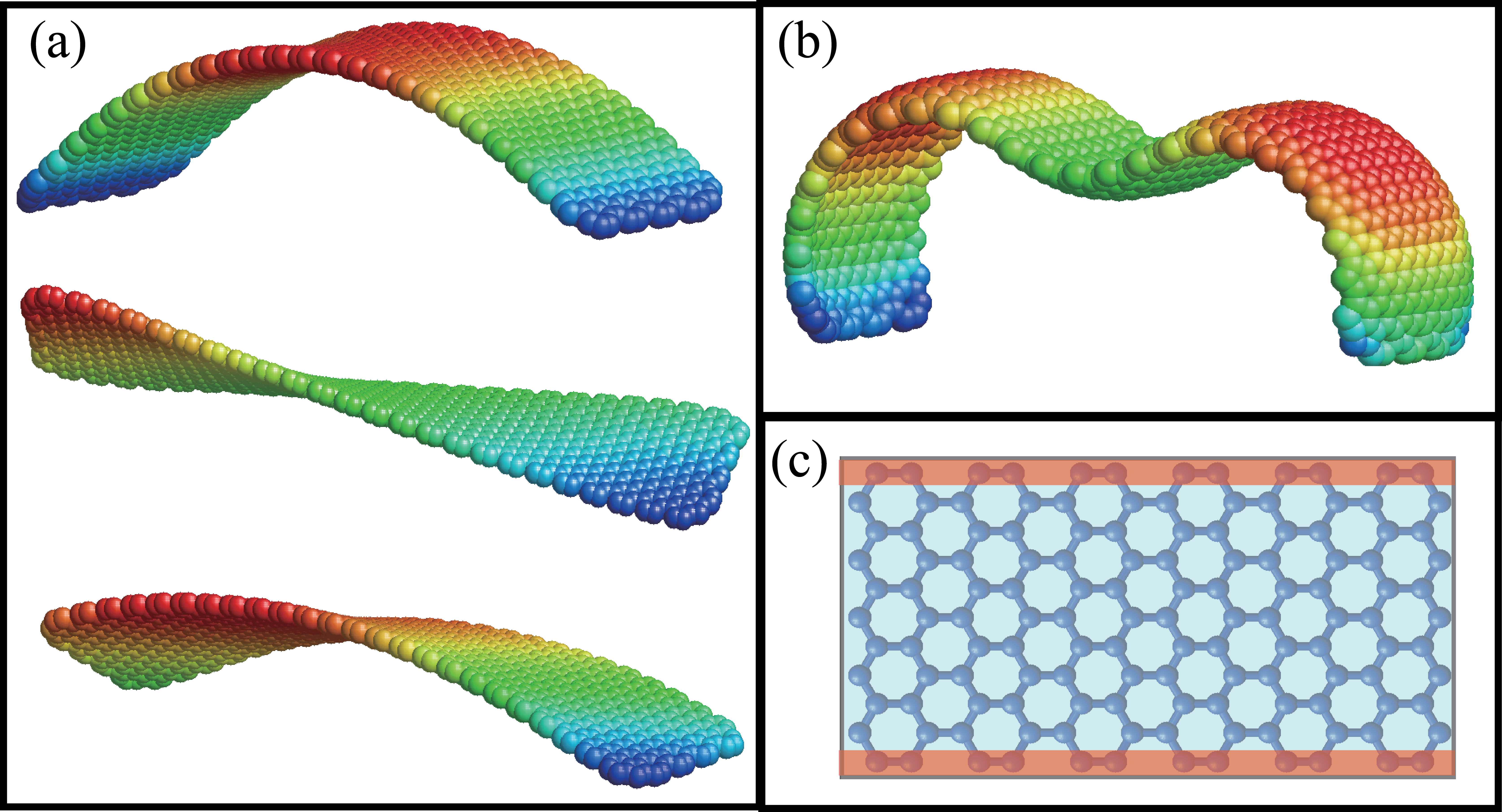}
\caption{\label{fig:fig1}  (color online) (a-b) Shapes observed in atomic-scale computations on 10.3\,nm$\times$2.2\,nm AGNRs with pristine ($\tau_e<0$) and reconstructed edges (usually $\tau_e>0$). (a) Saddle-like and twisted shapes for $\tau_e<0$. The two shapes also co-exist (bottom). (b) Curled shapes for $\tau_e>0$. The computations were performed using a reactive bond order (AIREBO) potential~\cite{pot:Stuart:2000,*md:Plimpton:1995}; see SI for details. 
(c) Schematic illustration of the core-edge composite framework.}
\end{figure}

The computed shapes are summarized in Fig.~\ref{fig:fig1}. In addition to the twisted and rippled shapes, we routinely observe transversely buckled saddle-like or curled shapes. 
As confirmation, a subset of these shapes have also been reported in past studies~\cite{gph:ShenoyZhang:2008, gnr:BetsYakobson:2009}. A recent study on curled ribbons (positive Gaussian curvature) with $\tau_e>0$ shows that the morphology is strongly width dependent as the ribbon core now deforms out-of-plane to accommodate the compressed edges~\cite{gph:ShenoyZhang:2010}. Interestingly, these shapes can also co-exist (Fig.~\ref{fig:fig1}a and Fig.~S1).

In order to quantify the relative stability, we employ a composite framework consisting of nanoribbon core, approximated as an infinite linear isotropic elastic thin plate (width $w$, and length $l$ $({h}\ll{w}\ll{l})$), glued onto ribbon edges approximated as elastic strings that stretch or compress in accordance with the edge stress (see Fig. \ref{fig:fig1}c). For long ribbons, the mid-surface traverse forces are negligible and the governing equation for the out-of-plane deflection $\zeta$ reduces to (see Supplementary Documents)~\cite{book:Mansfield:1989},
\begin{align}
\label{eq:Von1}
D(\zeta_{,xxxx}+2\zeta_{,xxyy}+\zeta_{,yyyy})
&=T\zeta_{,xx}\\
\label{eq:Von2}
-S(\zeta_{,xx}\zeta_{,yy}-\zeta_{,xy}\zeta_{xy}) &= T_{,xx}.
\end{align}
where $T$ is longitudinal force per unit length (line tension). 
In our framework, an additional incompatibility arises at the core-edge interface that relates the ribbon force $T$ to the edge stress $\tau_e$, 
\begin{equation}
\label{Compatibility}
\{{T/S}\}_{\pm{\frac{w}{2}}}=(\tau_e^{\pm}-\tau_e)/S_e,
\end{equation}
where we have set the origin at the midline and $\tau_e^\pm$ are the post-buckled edge stresses at the two edges.


To make contact with these atomically thin systems, it becomes necessary to seek solutions in the shallow-shell limit. A natural choice for the out-of-plane deflections is
\begin{equation}
\label{eq:Saddle}
\zeta(x,y)=-\frac{1}{2}\kappa x^2-{\theta}xy+f(y),
\end{equation}
where $\kappa$ is the longitudinal curvature and $\theta$ is the twist per unit length, both assumed constant along the ribbon length. The boundary value problem for the shape function $f(y)$ and the ribbon tension $T$ follows from substituting Eq.~\ref{eq:Saddle} into Eqs.~\ref{eq:Von1},~\ref{eq:Von2} together with free torques and shear force balance at the edges,
\begin{align}
\label{eq:Von_curve1}
f_{,yyyy}+\kappa\,\frac{T}{D}=0\; \text{and}\;\; T_{,yy}-S(\theta^2+\kappa\,f_{,yy})=0,\nonumber\\
\left\{f_{,yy}-\nu\kappa\right\}\big|_{\pm{\frac{w}{2}}}=0,\;\; \text{and}
\left\{f_{,yyy}\mp\kappa\,\frac{\tau_e^{\pm}}{D}\right\}\Big|_{\pm{\frac{w}{2}}}=0.
\end{align}
The solution that satisfies Eq.~\ref{eq:Von_curve1} is of the form
\begin{align}
\label{eq:Result}
& f=-\frac{\theta^2y^2}{2\kappa}+C_1\cosh{ky}\cos{ky}+C_2\sinh{ky}\sin{ky}+C_3\nonumber\\
& T=S\kappa\;(C_1\cosh{ky}\cos{ky}+C_2\sinh{ky}\sin{ky}).
\end{align}
and is more conveniently expressed in terms of scaled (dimensionless) variables: curvature $\bar{\kappa}\equiv\sqrt{S/D}w^2\kappa$, twist (per unit length) $\bar{\theta}\equiv\sqrt{S/D}w^2\theta$, edge stress $\bar{\tau}_e\equiv \tau_e w/D$ and edge modulus, $\bar{S}_e\equiv S_e/Sw$. In particular, the wavenumber $kw=\sqrt{\bar{\kappa}/2}$ and the constant $C_3$ are chosen such that the ribbon cannot translate, $\int^{w/2}_{-w/2}f(y)\,dy=0$ (Eq.~).
The balance between the energy stored in bending and stretching the ribbon forms the basis for the post-buckled shapes. Setting the energy of a flat relaxed ribbon as our reference, the excess energy $\delta\mathcal{E}$ (per unit ribbon length) follows from the shape function and ribbon tension (Eq.~S5 in SI) and has contributions from bending the core $\mathcal{E}_b=\mathcal{E}_b^c$ (both mean and Gaussian curvature) and stretching the core and the edge, $\mathcal{E}_s=\mathcal{E}_s^c + \mathcal{E}_s^e$.  Minimizing the ribbon elastic energy with respect to the curvature and twist 
yields the equilibrium shape.

\begin{figure*}[htb]
\includegraphics[width=1.7\columnwidth]{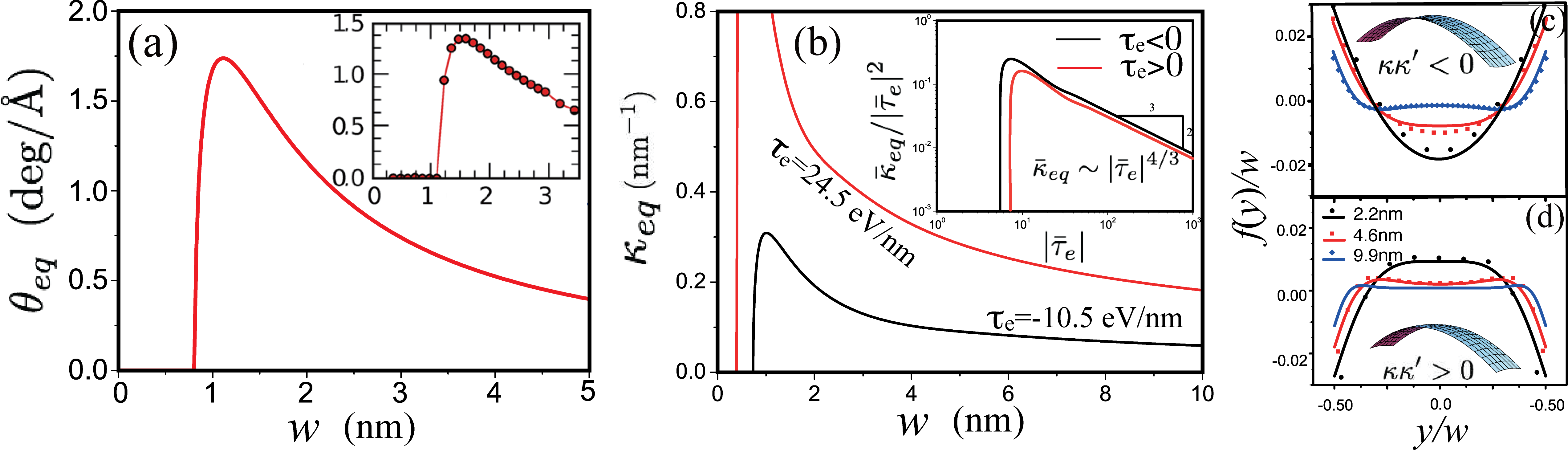}
\caption{\label{fig:fig2} (color online) Results based on our composite framework. (a) $\theta_{eq}$\,vs.\,$w$ for pristine AGNRs ($\tau_e=-10.5$\,eV/nm). (inset)~The same variation from quantum computations reported in Ref.~\cite{gnr:KitKoskinen:2010} and reproduced here. (b) Width dependence of longitude curvature $\kappa_{eq}$ for AGNRs with pristine and reconstructed edges, $\tau_e=-10.5$\,eV/nm and $\tau_e=24.5$\,eV/nm respectively. (inset) Scaled plot $\bar{\kappa}_{eq}/|\bar{\tau}_e|^2$\,vs.\,$|\tau_e|$ in the limit $S_e=0$ for compressive (black) and tensile (red) edge stresses. (c-d) Cross-sectional profiles $f(y)/w$\,vs\,$y/w$ for the AGNRs with widths ranging from $w=1-10$\,nm and (c) $\tau_e=-10.5$\,eV/nm and (d) $\tau_e=24.5$\,eV/nm. (insets) Schematic illustrations of the ribbon shapes.}
\end{figure*}
We first present results for pure twist mainly to benchmark our framework with respect to recent quantum calculations in AGNRs.  Setting 
$\kappa\rightarrow0$ in Eq.~\ref{eq:Result} results in a relatively simple expression for the equilibrium twist, 
\begin{equation}
\bar{\theta}_{eq}=0, \pm2\sqrt{30}\sqrt{\frac{-\bar{\tau}_e - 6(1-\nu)(1+2\bar{S}_e)}{(1+12\bar{S}_e)}}.
\end{equation}
It is immediately clear that there exists a critical compressive edge stress $|\bar{\tau}_e^*|= 6(1-\nu)(1+2\bar{S}_e)$ below which the ribbon is flat.
The underlying energetics is detailed in SI, but the relation can also be reasoned from a simple scaling analysis per unit ribbon length. Reverting to unscaled variables, the bending energy scales quadratically with the twist angle, $\mathcal{E}_b=D{\theta}^2{w}$. The decrease in stretching energy arises from the extension of the edge, $\mathcal{E}_s^e\sim -\tau_e^r\theta^2w^2$, where $\tau_e^r=\tau_e/(1+2\bar{S}_e)$ is the residual edge stress following in-plane relaxation. At the onset of buckling, $\mathcal{E}_b{\sim}\mathcal{E}_s^e$ which yields the critical edge stress, $\bar{\tau}_e^*\sim - (1+2\bar{S}_e)$. 
In the limit of large in-plane stiffness  $\bar{S}_e\ll1$ (e.g.~GNRs), the critical stress is approximately constant, $\bar{\tau}_e^\ast\approx - 6(1-\nu)$, i.e. the critical width below which the ribbon is flat scales inversely with the edge stress, $w^\ast\approx - 6(1-\nu)D/\tau_e$. Past $w^\ast$ the twist angle scales as $\bar{\theta}_{eq} \propto (\bar{\tau}_e-\bar{\tau}_e^\ast)^{1/2}$, characteristic of a supercritical pitchfork bifurcation. At large widths, the twist decreases {\it nonlinearly}, $\theta_{eq} \propto (-\tau_e/D)^{1/2}S^{-1/2}w^{-3/2}$ (Eq.~S8).

Figure~\ref{fig:fig1}a shows the width dependence for pristine AGNRs ($\nu=0.17$, $D=1.5$\,eV, $S=2000$\,eV/nm$^2$~\cite{gph:ShenoyZhang:2008}) with compressive edge stresses $\tau_e=-10.5$\,eV/nm and an edge stiffness $S_e=112.6$\,eV/nm~\cite{gph:ShenoyZhang:2008, gnr:ReddyShenoyZhang:2009}. The critical width is less than a nanometer, $w^\ast\approx0.8$\,nm and the twist exhibits a maximum, $\theta_{max}\approx1.75$\,deg/{\AA} at $w=1$\,nm (minimum pitch length $\lambda_{min}\approx20.6$\,nm), which separates the two distinct regimes of behavior described above. The inset in Fig.~\ref{fig:fig1}a  shows the width dependence reported in recent quantum computations on AGNRs. The overall trends are in excellent agreement with our results, especially at small widths, suggesting that, at least in the case of AGNRs, the edge-edge interactions are primarily elastic in nature. The small deviations (underestimation of the critical width and overestimation of the maximum twist $\theta_{max}\approx1.4$\,deg/{\AA} at $w\approx14$\,nm) are likely due to the differences computational frameworks (classical versus quantum). At large widths, the twist decays almost linearly in the computations; a similar dependence was also seen in separate atomistic computations~\cite{gnr:BetsYakobson:2009}. This is in contrast to the power law behavior and the discrepancy is primarily due to the narrow range of widths accessible to the computations.

We now turn our attention to pure bending. The solution for the equilibrium curvature $\kappa_{eq}$ again follows from Eq.~\ref{eq:Result} in the limit $\theta\rightarrow0$. The expressions are rather lengthy and for the sake of brevity, the width dependence is plotted in Fig.~\ref{fig:fig2}b for GNRs with varying edge stresses. The overall trend is similar to that for $\theta_{eq}(w)$ in that the flat ribbons spontaneously bend past a critical width $w^\ast$ with an equilibrium curvature $\kappa_{eq}$ that rapidly increases, exhibits a maximum and then decays non-linearly with at large widths. The critical (scaled) edge stress for onset of buckling is similar to that for twisting, $\bar{\tau}_e^*=\sqrt{5}(\sqrt{5}\nu\pm\sqrt{6-\nu^2})(1+2\bar{S}_e)$ (see SI for details) and is consistent with simple scaling analysis. For small edge deflections $\delta{\ll}w$, the transverse curvature is $\kappa^\prime\sim\delta/w^2$ and the curvature preserves the moment balance along the ribbon, $\kappa\sim\tau_e^r{\delta}/Dw$. Then, the bending energy (per unit length) scales as $\mathcal{E}_b=D(\kappa^2+\delta^2/w^4)w$. The decrease in energy is due to edge extension/compression and scales as $\mathcal{E}_e=-\tau_e^r\kappa\delta$. Equating $\mathcal{E}_b{\sim}\mathcal{E}_e$ yields the scaling, $|\bar{\tau}_e^*|\sim (1+ 2\bar{S}_e)$.  
The plot shows that the critical width for $\tau_e=-10.5$\,eV/nm is also of the same order as that for the twist, to within an angstrom. The quantitative similarities between the two shapes suggests a small energy difference; we address this in detail towards the end of this letter. 

The critical curve for $\tau_e=24.5$\,eV/nm plotted in Fig.~\ref{fig:fig2}b corresponds to reconstructed AGNRs and highlights the effect of edge stress. The equilibrium curvature in general is larger with a smaller critical width. However, the qualitative trends remain unchanged, 
suggesting a more general interplay between $\kappa_{eq}$ and $\tau_e$ during post-buckling. As confirmation, the inset in Fig.~\ref{fig:fig2}b shows the scaled plot, $\bar{\kappa}_{eq}/|\bar{\tau}_e|^2$ vs. $|\bar{\tau}_e|$, predicted by our framework in the limit $\bar{S}_e=0$.
Positive edge stresses result in curled ribbons with larger critical width -  the difference is entirely due to the Poisson's ratio (see SI). At large widths, $\bar{\kappa}_{eq}\propto|\bar{\tau}_e|^{n}$ with the scaled exponent $n=4/3$. The unscaled curvature therefore exhibits a power-law decay with width, $\kappa_{eq} \propto w^{-2/3}$,  consistent with recently reported results for GNRs with $\tau_e>0$. At intermediate widths in the vicinity of the peak curvature,  the exponent oscillates about $n=4/3$ (see SI). The deviations in the scaling are associated with transitions in the ribbon profile. Figure~\ref{fig:fig2}c-d shows these profiles for saddle-like and curled AGNR profiles as a function of width. Note the excellent agreement between our framework and computations. The profile for $w=9.9$\,nm is representative of the shapes at large widths in that the out-of-plane deflection are strongly localized at edges that do not interact. The edge boundary layer is preferentially strained and therefore bent, resulting in a  double-well profile with a  relatively flat midsection.  At small widths ($w=2.2$\,nm and $4.6$\,nm), the deflection penetrates through the width due to the edge-edge interactions mediated by a strained core as well as the boundary conditions. The profiles evolve such that the mean curvature $\kappa+\kappa^\prime$ decreases with width.

\begin{figure*}[htb]
\includegraphics[width=1.7\columnwidth]{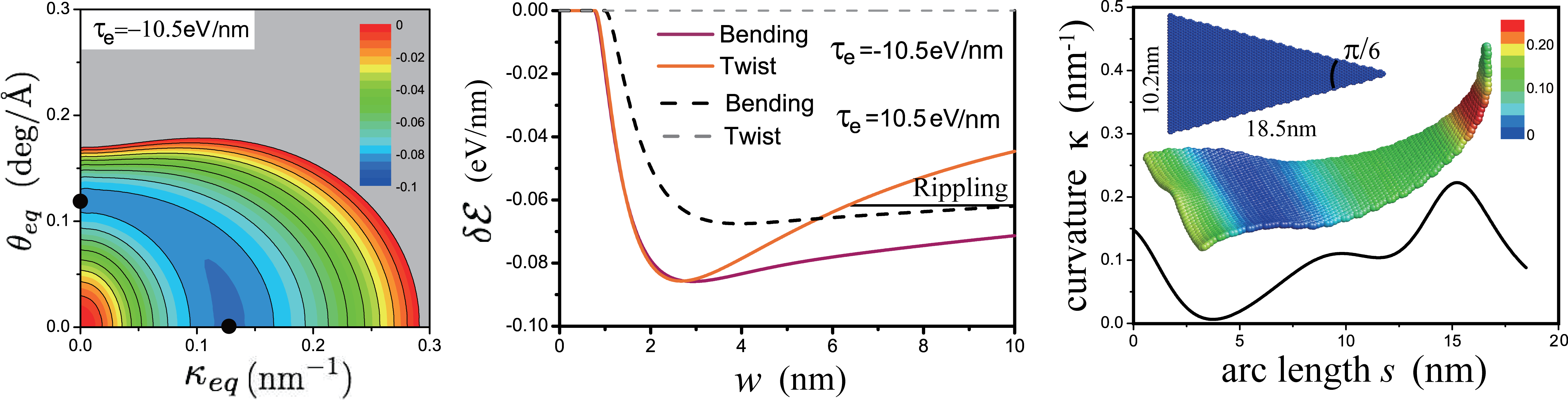}
\caption{\label{fig:fig3} (left) Contour plot of the elastic energy change $\delta \mathcal{E}$  predicted by our framework for bent and twisted AGNRs with pristine edges ($\tau_e=-10.5$\,eV/nm). (middle) Width dependence $\delta \mathcal{E}$(w) for minimal shapes of pristine and reconstructed AGNRs ($\tau_e=\pm10.5$\,eV/nm). The energy for edge rippling (solid black line) is based on wavelength observed in simulations and an assumed form for the shape~\cite{gph:ShenoyZhang:2008}. See text and SI for details. (c) (right, top) The minimal shape for an 18.5\,nm long tapering ribbon with opening angle is $\alpha=\pi/6$. The vicinal edges are at an angle $\pi/12$ off the armchair orientation. The color indicates the longitudinal curvature $\kappa$. (right, bottom) Variation in the ribbon curvature $\kappa$(s) towards the tapering end.}
\end{figure*}
Compressive edge stresses also lead to co-existing twisted and saddle-like shapes (Fig.~\ref{fig:fig1}a). 
Insight into the relative stability can be gleaned from the athermal behavior, readily available through our framework for any combination of $\theta$ and $\kappa$ (see SI). The results are plotted in Fig.~\ref{fig:fig3}a as a contour plot of the excess energy $\delta\mathcal{E}$ for AGNRs with $\tau_e=-10.5$\,eV/nm and fixed width, $w=3$\,nm. The colored regions represent energy gain associated with the morphology, i.e. $\delta\mathcal{E}<0$ with respect to planar yet relaxed ribbons. The origin ($\theta=\kappa=0$) is an unstable local maximum, as expected. The local minima for pure twist and bending (solid circles) correspond to the equilibrium points in Figs.~\ref{fig:fig2}a and~\ref{fig:fig2}b. Pure bending is favored, yet the energetic cost associated with a co-existing twist is exceedingly small ($0.01$\,eV/nm) and can be easily overcome by the thermal energy available at room temperatures ($k_BT\approx0.025$\,eV/atom). 

The width dependence of the energy gain associated with both morphological classes yields a comprehensive picture of the shape transitions. Figure~\ref{fig:fig3}b shows such a plot for GNRs with $\tau_e=\pm10.5$\,eV/nm. The individual contributions are plotted in detailed in SI. For compressive edge stress, the two shapes are energetically similar for $w^\ast<w<3$\,nm; the behavior is consistent with our observations so far. Pure bending becomes more favorable beyond a transition width $w=3$\,nm. In the large width limit, the energy gap that stabilizes the saddle-like morphologies scales as $\delta\mathcal{E} (\theta=0) - \delta\mathcal{E} (\kappa=0)$$\sim \tau_e^2/(Sw) (|\bar{\tau}_e|^{2/3} -1)$. The non-linear dependence on the edge stress offers insight into the relative shape stability in general; small compressive edge stresses (e.g.~hydrogenated GNRs) stabilize the twist such that the transition to saddle shapes shifts to larger ribbon widths, while large stresses favor saddle shapes.

At large widths, the edges will eventually ripple~\footnote{A similar transition was recently studied in the context of shapes of naturally occuring laminae such as long leaves, using a combination of linear stability analysis (rippling) and post-buckled solution (saddle shapes)~\cite{elastica:LiangMahadevan:2009}.}. While a general treatment of this transition is beyond the scope of the present study, the transition width for the specific case of AGNRs can be estimated based on the equilibrium wavelengths observed in prior computations ($\lambda=5.6$\,nm)~\cite{gnr:BetsYakobson:2009, gph:ShenoyZhang:2008} and an assumed shape of the ripples (see SI). Note that the corrections due to the edge-edge interactions are absent in our estimate. Based on the approximate excess energy of a ribbon so rippled (Fig.~\ref{fig:fig3}b), we expect a twist-rippling transition at $w=6$\,nm and a bending-rippling transition at widths of the order of tens of nanometer. The transition width should again scale inversely with the edge stress. For example, computations on AGNRs with $\tau_e\approx-26$\,eV/nm show that  edge ripples become favorable relative to twist (the saddle shapes were ignored) at widths as low as $1.5$\,nm~\cite{gnr:BetsYakobson:2009}. Finally, in the case of tensile edge stresses, the twisted morphology is obviously not viable. Interestingly, the energy associated with curled shapes is higher than that for saddle-shapes for the same width. At much larger widths, the compressed midline buckles into dimples that decay towards the edges, analogous to the bending-rippling transition. 

In conclusion, the interplay between ribbon width and mechanics that sets the shape highlights the importance of structural and chemical control of the edges. It also raises intriguing questions on shapes in ribbons with local variations in geometry and/or edge properties. As a case in point, Fig.~\ref{fig:fig3}c shows the computed minimal shape of a tapering GNR~\cite{gnr:SenNovoselovBuehler:2010}. The shape changes towards the tapering end in accordance with our predictions; the wider end is rippled, settles into a saddle-like curved shape, and then flattens out into a slightly twisted tapering end. The shape is not unlike the curled tips of growing leaf blades and flower petals, highlighting a general principle where geometrical (and possibly material) modulations are naturally amplified as controllable shape transitions.


%
\end{document}